\documentclass[traditabstract]{aa} 

\usepackage{graphicx}
\usepackage{txfonts}
\usepackage{natbib}

\begin{document}

\title{Both accurate and precise $gf$-values for \ion{Fe}{ii} lines
\thanks{Based in part on observations obtained at the W. M. Keck Observatory, which is operated 
jointly by the California Institute of Technology, the University of California, 
and the National Aeronautics and Space Administration.}
}
\titlerunning{Oscillator Strengths for \ion{Fe}{ii} lines}

\newcommand{\teff}{T$_{\rm eff}$ }
\newcommand{\tsin}{T$_{\rm eff}$}

\author{
J. Mel\'endez\inst{1} \and
B. Barbuy\inst{2}
}


\institute{
Centro de Astrof\'{\i}sica da Universidade do Porto, Rua das Estrelas, 4150-762 Porto, Portugal  \and
Universidade de S\~{a}o Paulo, IAG, Rua do Mat\~{a}o 1226, 
Cidade Universit\'{a}ria, S\~{a}o Paulo 05508-900, Brazil}

\date{Received ...; accepted ...}

\abstract{
We present a new set of oscillator strengths for 142 \ion{Fe}{ii} lines in the
wavelength range 4000-8000 \AA.
Our $gf$-values are both accurate and precise, because each multiplet was 
globally normalized using laboratory data (accuracy), while the relative $gf$-values
of individual lines within a given multiplet were obtained from 
theoretical calculations (precision). Our line list was tested
with the Sun and high-resolution ($R \approx 10^5$), high-S/N ($\approx$ 700-900) 
Keck+HIRES spectra of the metal-poor stars HD 148816 and HD 140283, for which
line-to-line scatter ($\sigma$) in the iron abundances from \ion{Fe}{ii} lines
as low as 0.03, 0.04, and 0.05 dex are found, respectively.
For these three stars the standard error in the mean
iron abundance from \ion{Fe}{ii} lines is negligible ($\sigma_\mathrm{mean} \leq$ 0.01 dex).
The mean solar iron abundance obtained using our $gf$-values and
different model atmospheres is $A_{Fe}$ = 7.45 ($\sigma$ = 0.02).
}


\maketitle

%

\section{Introduction}

The iron abundance determined from \ion{Fe}{ii} lines is more reliable 
than that obtained from \ion{Fe}{i} lines, as  \ion{Fe}{ii} depends little on the 
details in the temperature structure of model
atmospheres and it is almost immune to departures from LTE
(e.g. Th\'evenin \& Idiart 1999; Shchukina \& Trujillo Bueno 2001; Gehren et al. 2001; Asplund 2005).
Although some authors argue that small departures from LTE
may be present for \ion{Fe}{ii} (e.g. Shchukina, Trujillo Bueno \& Asplund 2005;
Mel\'endez et al. 2006a), the effects are much less for \ion{Fe}{ii} 
than for \ion{Fe}{i}. Thus, \ion{Fe}{ii} has recently been
used as the preferred indicator for iron abundances in F-G-K type stars
(e.g. Nissen et al. 2002, 2004, 2007; Mel\'endez \& Barbuy 2002;
Kraft \& Ivans 2003; Asplund et al. 2006; Mel\'endez et al. 2006a; Ramirez et al. 2006),
although some problems may be present 
for metal-rich late K dwarfs (e.g. Yong et al. 2004;
Ramirez, Allende Prieto \& Lambert 2007).

The robustness of \ion{Fe}{ii} is undermined by the
uncertainty in their $gf$-values. 
As is well known in the literature, there is a lack of precise
transition probabilities for \ion{Fe}{ii} lines, so that even the best
available laboratory data introduce large uncertainties (at the level of 0.1 dex) 
in the determination of iron abundances (see e.g. Grevesse \& Sauval 1999).
New laboratory experiments and theoretical calculations 
(see Fuhr \& Wiese 2006 and references therein) have not improved
the situation, as will be shown in Sects. 3-4.

During the last few years we have critically evaluated each 
\ion{Fe}{ii} multiplet (as first described in Mel\'endez \& Barbuy 2002), 
in order to improve the precision of the available data. 
Our whole line list has never been fully described or published, yet
it is already being widely used in the literature (Mel\'endez \& Barbuy 2002;
Barbuy et al. 2006, 2007; Coelho et al. 2005; Mel\'endez et al. 2006a, 2006b;
Alves-Brito et al. 2005, 2006; Zoccali et al. 2004; Smiljanic et al. 2006, 2008;
Allen \& Barbuy 2006; Ramirez et al. 2006; 
Ramirez, Allende Prieto \& Lambert 2007, Hekker \& Mel\'endez 2007; 
Santos et al. 2009). In the present 
paper we present our improved oscillator strengths for \ion{Fe}{ii} lines,
and we show that they are very precise and accurate and should be
adopted until better laboratory and theoretical data are available.

\begin{table*}
\begin{minipage}[t]{\textwidth}
\caption{Fe II line list.}
\label{fe2list}
\centering          
\renewcommand{\footnoterule}{}  
\begin{tabular}{cccccccccccccccccc} 
\hline\hline                
\scriptsize
{$\lambda$} & {$\chi_{exc}$} & {log {\it gf}} & $C_6$ & {$\lambda$} & {$\chi_{exc}$} & {log {\it gf}} & $C_6$ & {$\lambda$} & {$\chi_{exc}$} & {log {\it gf}} & $C_6$\\
{({\rm \AA})}   & {(eV)}  & {}  & {} & {({\rm \AA})}   & {(eV)}  & {}  & {} & {({\rm \AA})}   & {(eV)}  & {}  & {} \\
\hline    
 4087.284 & 2.5828 &-4.57$^L$ & (0.857E-32)& 4855.554 & 2.7045 &-4.46$^S$ & 0.787E-32  & 6147.741 & 3.8887 &-2.69$^S$ & 0.943E-32  \\ 
 4122.668 & 2.5828 &-3.26$^L$ & 0.869E-32  & 4871.277 & 2.7045 &-4.25$^S$ & 0.787E-32  & 6149.258 & 3.8894 &-2.69$^S$ & 0.943E-32  \\ 
 4128.748 & 2.5828 &-3.63$^L$ & 0.956E-32  & 4893.820 & 2.8283 &-4.21$^S$ & 0.787E-32  & 6150.098 & 3.2215 &-4.73$^S$ & 0.787E-32  \\ 
 4173.461 & 2.5828 &-2.65$^L$ & 0.943E-32  & 4923.927 & 2.8912 &-1.26$^L$ & 0.810E-32  & 6179.384 & 5.5687 &-2.62$^S$ & 0.155E-31  \\ 
 4178.862 & 2.5828 &-2.51$^L$ & 0.857E-32  & 4924.921 & 2.8443 &-4.90$^S$ & 0.787E-32  & 6184.929 & 5.5709 &-3.72$^S$ & 0.787E-32  \\ 
 4233.172 & 2.5828 &-1.97$^L$ & 0.943E-32  & 4991.126 & 2.7786 &-4.55$^S$ & 0.787E-32  & 6233.534 & 5.4845 &-2.51$^S$ & 0.155E-31  \\ 
 4258.154 & 2.7045 &-3.33$^L$ & 0.869E-32  & 4993.358 & 2.8066 &-3.62$^S$ & 0.775E-32  & 6238.392 & 3.8887 &-2.60$^L$ & 0.943E-32  \\ 
 4273.326 & 2.7045 &-3.38$^L$ & 0.956E-32  & 5000.743 & 2.7786 &-4.61$^S$ & 0.787E-32  & 6239.953 & 3.8894 &-3.41$^S$ & 0.943E-32  \\ 
 4278.159 & 2.6921 &-3.73$^L$ & (0.845E-32)& 5018.440 & 2.8912 &-1.10$^L$ & 0.798E-32  & 6247.350 & 6.2090 &-1.98$^S$ & 0.881E-32  \\ 
 4296.572 & 2.7045 &-2.92$^L$ & 0.869E-32  & 5036.920 & 2.8283 &-4.67$^S$ & 0.775E-32  & 6247.557 & 3.8918 &-2.30$^S$ & 0.943E-32  \\ 
 4303.176 & 2.7045 &-2.56$^L$ & 0.943E-32  & 5100.664 & 2.8066 &-4.17$^S$ & 0.787E-32  & 6248.898 & 5.5110 &-2.67$^S$ & 0.159E-31  \\ 
 4351.769 & 2.7045 &-2.25$^L$ & 0.943E-32  & 5120.352 & 2.8283 &-4.24$^S$ & 0.787E-32  & 6317.983 & 5.5110 &-1.96$^S$ & 0.159E-31  \\ 
 4369.411 & 2.7786 &-3.65$^L$ & 0.869E-32  & 5132.669 & 2.8066 &-4.08$^S$ & 0.775E-32  & 6331.954 & 6.2173 &-1.88$^S$ & 0.881E-32  \\ 
 4384.319 & 2.6570 &-3.44$^L$ & 0.845E-32  & 5136.802 & 2.8443 &-4.43$^S$ & 0.787E-32  & 6369.462 & 2.8912 &-4.11$^L$ & 0.742E-32  \\ 
 4385.387 & 2.7786 &-2.66$^L$ & 0.943E-32  & 5146.127 & 2.8283 &-3.91$^S$ & 0.787E-32  & 6371.125 & 5.5491 &-3.13$^S$ & 0.153E-31  \\ 
 4413.601 & 2.6759 &-3.79$^L$ & 0.845E-32  & 5150.941 & 2.8557 &-4.48$^S$ & 0.787E-32  & 6383.722 & 5.5526 &-2.24$^S$ & 0.159E-31  \\ 
 4416.830 & 2.7786 &-2.65$^L$ & 0.943E-32  & 5154.409 & 2.8443 &-4.13$^S$ & 0.787E-32  & 6385.451 & 5.5526 &-2.59$^S$ & 0.159E-31  \\ 
 4472.929 & 2.8443 &-3.36$^L$ & (0.869E-32)& 5161.184 & 2.8557 &-4.47$^S$ & 0.787E-32  & 6416.919 & 3.8918 &-2.64$^S$ & 0.930E-32  \\ 
 4489.183 & 2.8283 &-2.96$^L$ & 0.869E-32  & 5169.033 & 2.8912 &-1.00$^L$ & 0.798E-32  & 6432.680 & 2.8912 &-3.57$^L$ & 0.742E-32  \\ 
 4491.405 & 2.8557 &-2.71$^L$ & 0.869E-32  & 5171.640 & 2.8066 &-4.54$^S$ & 0.775E-32  & 6433.814 & 6.2191 &-2.37$^S$ & 0.881E-32  \\ 
 4508.288 & 2.8557 &-2.44$^L$ & 0.956E-32  & 5197.577 & 3.2306 &-2.22$^L$ & 0.869E-32  & 6442.955 & 5.5491 &-2.44$^S$ & 0.155E-31  \\ 
 4515.339 & 2.8443 &-2.60$^L$ & 0.869E-32  & 5234.625 & 3.2215 &-2.18$^L$ & 0.869E-32  & 6446.410 & 6.2225 &-1.97$^S$ & 0.881E-32  \\ 
 4520.224 & 2.8068 &-2.65$^L$ & 0.857E-32  & 5238.624 & 2.8912 &-5.11$^S$ & (0.798E-32)& 6455.837 & 5.5526 &-2.92$^S$ & 0.159E-31  \\   
 4522.634 & 2.8443 &-2.25$^L$ & 0.943E-32  & 5256.938 & 2.8912 &-4.06$^S$ & 0.798E-32  & 6456.383 & 3.9036 &-2.05$^S$ & 0.930E-32  \\   
 4534.168 & 2.8557 &-3.28$^L$ & 0.869E-32  & 5264.812 & 3.2304 &-3.13$^L$ & 0.943E-32  & 6482.204 & 6.2191 &-1.78$^S$ & 0.881E-32  \\   
 4541.524 & 2.8557 &-2.98$^L$ & 0.943E-32  & 5276.002 & 3.1996 &-2.01$^L$ & 0.857E-32  & 6491.246 & 5.5851 &-2.76$^S$ & 0.160E-31  \\   
 4549.192 & 5.9113 &-1.62$^L$ & 0.142E-31  & 5284.109 & 2.8912 &-3.11$^S$ & 0.798E-32  & 6493.035 & 5.5851 &-2.55$^S$ & 0.160E-31  \\   
 4549.474 & 2.8283 &-2.09$^L$ & 0.943E-32  & 5316.615 & 3.1529 &-1.87$^L$ & 0.845E-32  & 6506.333 & 5.5895 &-2.68$^S$ & 0.159E-31  \\   
 4555.893 & 2.8283 &-2.40$^L$ & 0.857E-32  & 5316.784 & 3.2215 &-2.74$^L$ & 0.943E-32  & 6508.129 & 5.5895 &-3.45$^S$ & 0.159E-31  \\   
 4576.340 & 2.8443 &-2.95$^L$ & 0.943E-32  & 5325.553 & 3.2215 &-3.16$^L$ & 0.857E-32  & 6516.080 & 2.8912 &-3.31$^L$ & 0.742E-32  \\   
 4582.835 & 2.8443 &-3.18$^L$ & 0.857E-32  & 5337.732 & 3.2304 &-3.72$^L$ & 0.943E-32  & 6517.018 & 5.5851 &-2.73$^S$ & 0.159E-31  \\   
 4583.837 & 2.8068 &-1.93$^L$ & 0.930E-32  & 5362.869 & 3.1996 &-2.57$^L$ & 0.930E-32  & 6562.200 & 5.6052 &-2.83$^S$ & 0.160E-31  \\   
 4601.378 & 2.8912 &-4.48$^L$ & 0.918E-32  & 5414.073 & 3.2215 &-3.58$^L$ & 0.930E-32  & 6586.699 & 5.6052 &-2.74$^S$ & 0.160E-31  \\   
 4620.521 & 2.8283 &-3.21$^L$ & 0.930E-32  & 5425.257 & 3.1996 &-3.22$^L$ & 0.845E-32  & 6598.301 & 5.6156 &-3.05$^S$ & 0.157E-31  \\   
 4625.893 & 5.9560 &-2.35$^L$ & 0.143E-31  & 5432.967 & 3.2675 &-3.38$^S$ & 0.857E-32  & 7222.394 & 3.8889 &-3.26$^L$ & 0.956E-32  \\   
 4629.339 & 2.8068 &-2.34$^L$ & 0.845E-32  & 5525.125 & 3.2676 &-3.97$^L$ & 0.918E-32  & 7224.487 & 3.8891 &-3.20$^L$ & 0.956E-32  \\   
 4635.316 & 5.9560 &-1.42$^L$ & 0.143E-31  & 5534.847 & 3.2449 &-2.75$^S$ & 0.845E-32  & 7301.560 & 3.8916 &-3.63$^S$ & 0.857E-32  \\   
 4648.944 & 2.5828 &-4.58$^S$ & 0.775E-32  & 5591.368 & 3.2675 &-4.44$^S$ & 0.845E-32  & 7308.073 & 3.8889 &-3.03$^L$ & 0.943E-32  \\   
 4656.981 & 2.8912 &-3.60$^L$ & 0.918E-32  & 5627.497 & 3.3866 &-4.10$^L$ & 0.869E-32  & 7310.216 & 3.8891 &-3.37$^L$ & 0.943E-32  \\   
 4666.758 & 2.8283 &-3.28$^L$ & 0.845E-32  & 5657.935 & 3.4247 &-4.03$^L$ & 0.869E-32  & 7320.654 & 3.8918 &-3.23$^L$ & 0.943E-32  \\   
 4670.182 & 2.5828 &-4.09$^S$ & 0.775E-32  & 5725.963 & 3.4247 &-4.76$^L$ & 0.869E-32  & 7449.335 & 3.8889 &-3.27$^L$ & 0.943E-32  \\   
 4720.149 & 3.1974 &-4.48$^S$ & 0.930E-32  & 5732.724 & 3.3866 &-4.60$^L$ & 0.857E-32  & 7462.407 & 3.8918 &-2.74$^L$ & 0.943E-32  \\   
 4731.453 & 2.8912 &-3.10$^L$ & 0.905E-32  & 5813.677 & 5.5706 &-2.51$^L$ & 0.798E-32  & 7479.693 & 3.8916 &-3.61$^S$ & 0.857E-32  \\   
 4825.736 & 2.6350 &-4.87$^S$ & 0.775E-32  & 5991.376 & 3.1529 &-3.54$^S$ & 0.775E-32  & 7515.832 & 3.9036 &-3.39$^L$ & 0.943E-32  \\   
 4831.126 & 3.3394 &-4.89$^S$ & 0.943E-32  & 6084.111 & 3.1996 &-3.79$^S$ & 0.787E-32  & 7655.488 & 3.8918 &-3.56$^L$ & 0.930E-32  \\   
 4833.197 & 2.6572 &-4.64$^S$ & 0.775E-32  & 6113.322 & 3.2215 &-4.14$^S$ & 0.787E-32  & 7711.724 & 3.9034 &-2.50$^L$ & 0.930E-32  \\   
 4833.865 & 2.8443 &-5.11$^S$ & (0.787E-32)& 6116.057 & 3.2306 &-4.67$^S$ & 0.787E-32  &          &        &         &   \\   
 4839.998 & 2.6757 &-4.75$^S$ & 0.787E-32  & 6129.703 & 3.1996 &-4.64$^S$ & 0.775E-32  &          &        &         &   \\   
\hline                                 
\hline                                 
\end{tabular}                
\footnotetext{Note.- Lines from multiplets normalized using laboratory data are labelled $L$, and the Sun $S$.
The broadening constants $C_6$ are based on cross-section calculations by Barklem \& Aspelund-Johansson (2005).}\end{minipage}
\end{table*}

\section{Improved oscillator strengths}

Even though the bulk of laboratory $gf$-values are probably correct 
on an absolute scale; i.e., they are probably accurate, the oscillator 
strengths have large uncertainties on a line-by-line basis, meaning that they
are imprecise. On the other hand, theoretical calculations are 
not always correct on an absolute scale, but the theoretical relative line 
ratios within multiplets are reliable, except probably for 
lines with low $f$-values;
as indicated by our tests using \ion{Fe}{i} and \ion{Fe}{ii} lines,
the relative agreement between theoretical
and laboratory $gf$-values worsens for decreasing line strength
(note that this behaviour has also been noted by other authors, e.g. 
Goldbach, Martin \& Nollez 1989; Pickering, Johansson \& Smith 2001),
probably due to the difficulties in computing reliably $gf$-values for these
lines (see e.g. Bi\'emont et al. 1991; Raassen \& Uylings 1998).
In Mel\'endez \& Barbuy (2002) we
exploited the advantages of both laboratory and theoretical methods, 
adopting  relative line ratios within a given multiplet 
from theoretical calculations, whereas the absolute transition
probabilities for each multiplet were determined from laboratory measurements.
For the present work we revised the $gf$-values of \ion{Fe}{ii} lines,
using new laboratory and theoretical data.

We have adopted theoretical $gf$-values by Bi\'emont et al. (1991),
Raassen \& Uylings (1998), and recent calculations
by R. L. Kurucz\footnote{as published online in October 2003 and 
August 2008 at http://kurucz.harvard.edu}. Those data were calibrated with 
the following experimental data: lifetimes of upper levels from
Schnabel, Schultz-Johanning \& Kock (2004), 
Schnabel, Kock, \& Holweger (1999), Schnabel \& Kock (2001), 
Guo et al. (1992), Hannaford et al. (1992), and Bi\'emont et al.
(1991), and branching fractions by Heise \& Kock (1990), Pauls,
Grevesse, \& Huber (1990), and Kroll \& Kock (1987). The (absorption)
oscillator strength $f_{lu}$ is related to the branching fraction ($BF_{lu}$)
and the lifetime ($\tau_u$) by (e.g. Hannaford 1994):

\begin{equation}
 f_{lu} = 1.499 \times 10^{-7} (g_u/g_l) (BF_{lu}/\tau_u[ns]) \lambda^2[\AA]
\end{equation}

\noindent where $\tau_u$ is given in $ns$ and 
the wavelength of the transition $\lambda$ in \AA; 
$l$ and $u$ represent the lower and upper levels, respectively; 
$g_l$ and $g_u$ are the statistical weights of the lower and upper levels,
respectively, which depend on the total angular momentum of the level, i.e. on the 
quantum number $J$:

$$g = 2 J + 1.$$

\noindent Since the ratio $BF$/$\tau$ is equivalent to the transition probability $A$
($= BF/\tau$), the oscillator strength can also be obtained from

\begin{equation}
f_{lu} = 1.499 \times 10^{-7} (g_u/g_l) A_{ul}[10^9 s^{-1}] \lambda^2[\AA]
\end{equation}

\noindent where $A_{ul}$ is given in $10^9$ $s^{-1}$ and $\lambda$ in \AA.

When no laboratory measurement for any line of a multiplet was available,
the relative oscillator strengths were derived from theoretical
calculations, but the absolute $gf$-values of the multiplet were obtained 
from an inverse analysis based on the 
National Solar Observatory FTS solar flux spectrum by 
Hinkle et al. (2000), which is essentially the same spectrum as was previously 
published by Kurucz et al. (1984) but corrected for telluric absorption. 
The solar analysis was performed with the codes ABON 2002 (Spite 1967) and MOOG 2002 (Sneden 1973), 
using a spatially and temporally averaged 3D solar model atmosphere (hereafter
$<$3D$>$; Asplund et al. 2004) and
adopting a solar abundance obtained with the $<$3D$>$ model 
and the previously determined laboratory $gf$-values of the \ion{Fe}{ii} lines
and with interaction constants $C_6$ computed from broadening cross-sections $\sigma$
calculated by Barklem \& Aspelund-Johansson (2005), using the following relation
(derived from Gray (2005), as described in Coelho et al. (2005)):

\begin{equation}
C_6[cm^6 s^{-1}] = 6.46 \times 10^{-34} (\sigma[a.u.]/63.65)^{5/2}
\end{equation}

\noindent where the cross-section $\sigma$ is given in atomic units and
the interaction constant $C_6$ in $cm^6 s^{-1}$.

When the relative line ratios predicted by the 
theoretical calculations fail to reproduce the observed solar line ratios
within a given multiplet, we preferred to adopt $gf$-values based entirely
on laboratory measurements or to make slight corrections (usually no larger than 0.1 dex) 
to reproduce the solar spectrum better. This was mainly the
case for weak lines, because their theoretical line ratios may be incorrect.

The complete line list of 142 \ion{Fe}{ii} lines is given in Table 1,
where $gf$-values based on laboratory or solar measurements 
are labelled $L$ or $S$, respectively. The $C_6$ constants computed from
Eq. (3) are also given in the Table 1. For five lines (given
between parenthesis in Table 1) cross-sections
were not explicitly computed by Barklem \& Aspelund-Johansson (2005).
In those cases we obtained $C_6$ from other lines of the same
multiplet.

The procedure adopted here is very time-consuming, as each
multiplet is evaluated individually, but the results are worthwhile
and will significantly improve the precision of iron abundances 
obtained from \ion{Fe}{ii} lines, as shown in the next sections.

\section{The solar iron abundance}

As discussed by Grevesse \& Sauval (1999), it is very disappointing
that accurate transition probabilities are known for only a very
few \ion{Fe}{ii} lines, making the iron abundance obtained 
from \ion{Fe}{ii} very uncertain, with a line-to-line scatter
as high as 0.1 dex. 

To test that our $gf$-values reduce the scatter in
the solar iron abundance, we rescaled a 
few representative solar iron abundance determinations from the literature
using a total of 43 different \ion{Fe}{ii} lines 
(Bi\'emont et al. 1991; Hannaford et al. 1992; 
Schnabel et al. 1999; Asplund et al. 2000).
Since the scatter in the abundances is very high, we have used 
the median instead of the average here 
(and in the next section).
The results are shown in Table 2 and Figures 1-4. We succeed in all
cases to lower the uncertainties significantly, with an improvement
as high as almost a factor of 3 for the Hannaford et al. (1992) data (Fig. 1),
for which the $\sigma$ was reduced from 0.084 dex to only 0.030 dex.
The $-$0.05 dex correction proposed by Bi\'emont et al.
to their theoretical $gf$-values may not be necessary,
otherwise their mean solar iron abundance would be $A_{Fe}$ = 7.54.

The $gf$-values given by Schnabel et al. (2004) for lines used
in the solar iron abundance determination are very similar
to those given in Schnabel et al. (1999), and indeed the result
given by Schnabel et al. (2004) is the same (both in the
mean value and line-to-line scatter) as in
Schnabel et al. (1999), $A_{Fe}$ = 7.42 ($\sigma$ = 0.09); 
therefore, the revised values given by Schnabel et al. (2004) do not
improve the precision of their older transition probabilities.
The $gf$-values given in the critical compilation of Fuhr \& Wiese (2006)
in the optical region are mainly based on the laboratory results
of Schnabel et al. (2004), which are not very precise as discussed above,
and the (uncorrected) theoretical results of Raassen \& Uylings (1998), which 
are not very accurate. Indeed, the solar iron abundance obtained by
Raassen \& Uylings (1998) is 7.59, much higher than the meteoritic
iron abundance (7.45$\pm$0.03; Asplund, Grevesse \& Sauval 2005).

Given the importance of the solar iron abundance as a standard reference
in astronomy, we have computed its abundance using our $gf$-values for
the 15 lines analysed in Hannaford et al. (1992). The (centre of the disk)
equivalent widths adopted here are shown in Table 3; they are the average of the 
measurements presented in Holweger et al. (1990),
Bi\'emont et al. (1991), Hannaford et al. (1992), and Grevesse \& Sauval (1999).
The calculations were performed with MOOG 2002 using six model atmospheres: 
the Holweger \& M\"uller (1974) model atmosphere, Kurucz overshooting
(Castelli et al. 1997) and the latest no-overshooting (Castelli \& Kurucz 2004)
models, the MARCS 1997 (Asplund et al. 1997) and MARCS 2008 (Gustafsson et al. 2008)
models, and the $<$3D$>$ solar model (Asplund et al. 2004). The results
are shown in Table 4, where rescaled results based on a three-dimensional 
hydrodynamical model of the solar atmosphere (Asplund et al. 2000) are also presented.
The individual iron abundances for each model atmosphere have a very low line-to-line
scatter ($\sigma <$ 0.03 dex) and the 
standard error ($\sigma_{\rm mean} = \sigma/\sqrt{n}$) for the mean value
is negligible ($\sigma_{\rm mean} < 0.01$ dex). 
The different models give about the same result,
$A_{Fe} \approx$ 7.45 ($\sigma$ = 0.02),
but values as low as 7.42-7.43 (e.g. see Fig. 5 for the MARCS 2008 model), 
i.e., slightly lower than the
meteoritic iron abundance, are not unlikely,
although alternatively it may imply that the transformation between
the solar and meteoritic abundance scales may be slightly in error.
Interestingly, a recent work by
Mel\'endez et al. (2009), in a comparison of the Sun with solar twins, shows that 
the solar photosphere may be 
slightly depleted in refractory elements (including iron).

\begin{figure}
\resizebox{\hsize}{!}{\includegraphics{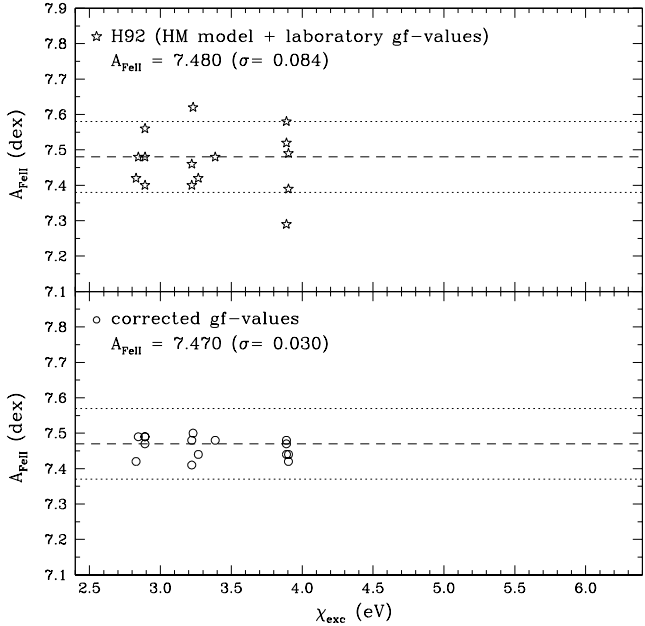}}
\caption{Iron abundances from \ion{Fe}{ii} lines obtained by Hannaford et al. (1992) using
their laboratory $gf$-values (upper panel) and re-scaled abundances using our improved
$gf$-values (lower panel). Dashed and dotted lines are shown at the median value and 
$\pm$ 0.1 dex, respectively.}                      
\label{h92}    
\end{figure}

\begin{figure}
\resizebox{\hsize}{!}{\includegraphics{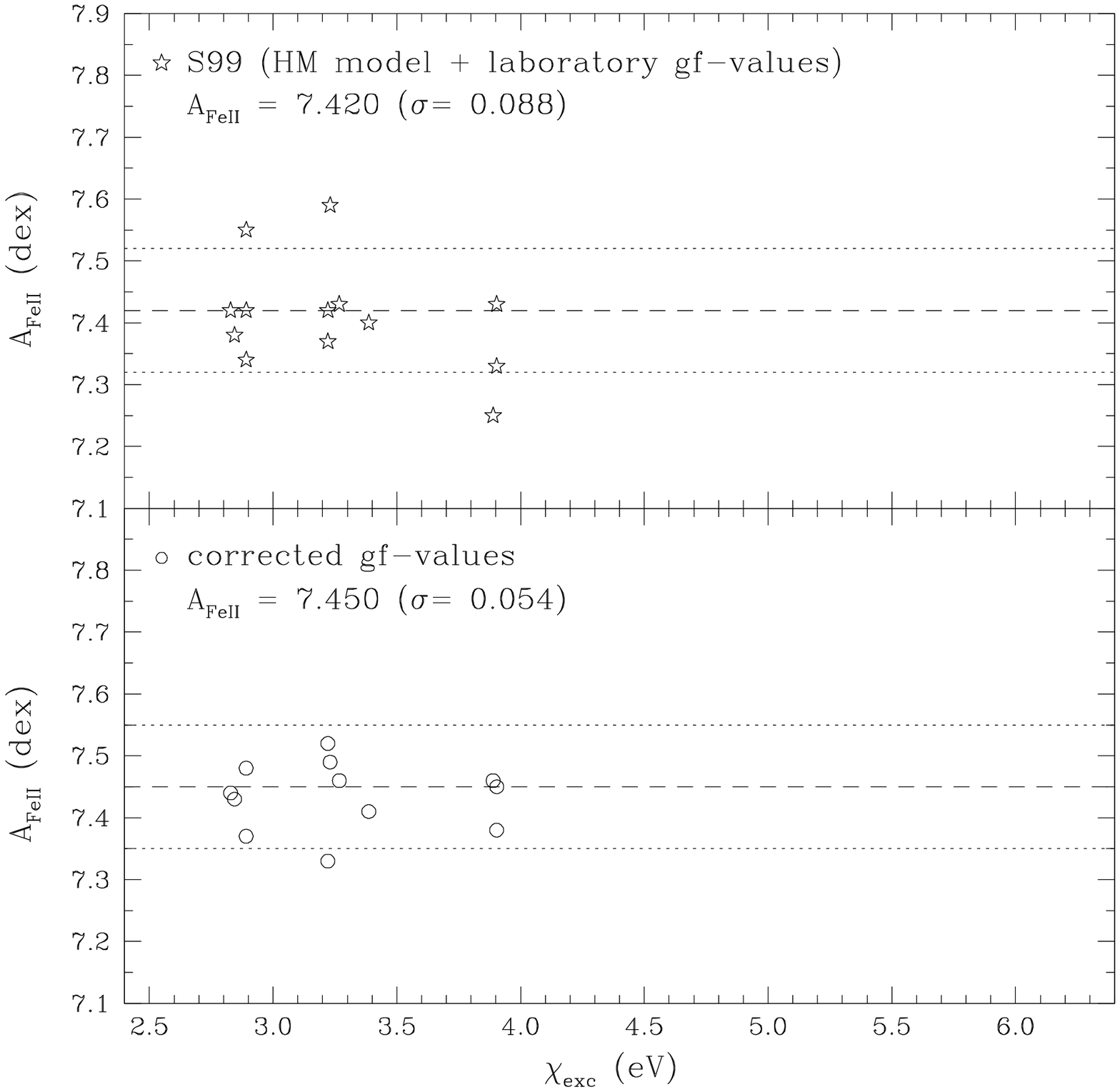}}
\caption{Iron abundances from \ion{Fe}{ii} lines obtained by Schnabel et al. (1999) using
their laboratory $gf$-values (upper panel) and re-scaled abundances using our improved
$gf$-values (lower panel). Dashed and dotted lines are shown at the median value and 
$\pm$ 0.1 dex, respectively.}                      
\label{s99}    
\end{figure}

\begin{figure}
\resizebox{\hsize}{!}{\includegraphics{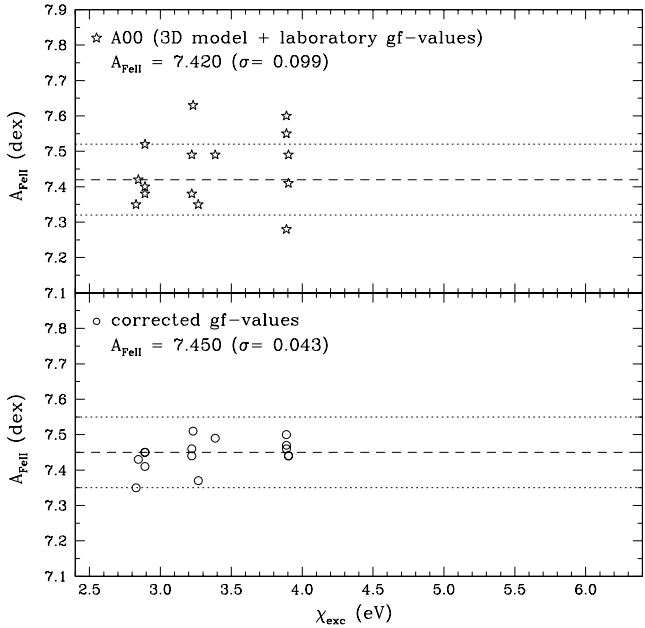}}
\caption{Iron abundances from \ion{Fe}{ii} lines obtained by Asplund et al. (2000) using
the Hannaford et al. (1992) laboratory $gf$-values (upper panel) and re-scaled abundances using our improved
$gf$-values (lower panel). Dashed and dotted lines are shown at the median value and 
$\pm$ 0.1 dex, respectively.}                      
\label{asplund}    
\end{figure}

\begin{figure}
\resizebox{\hsize}{!}{\includegraphics{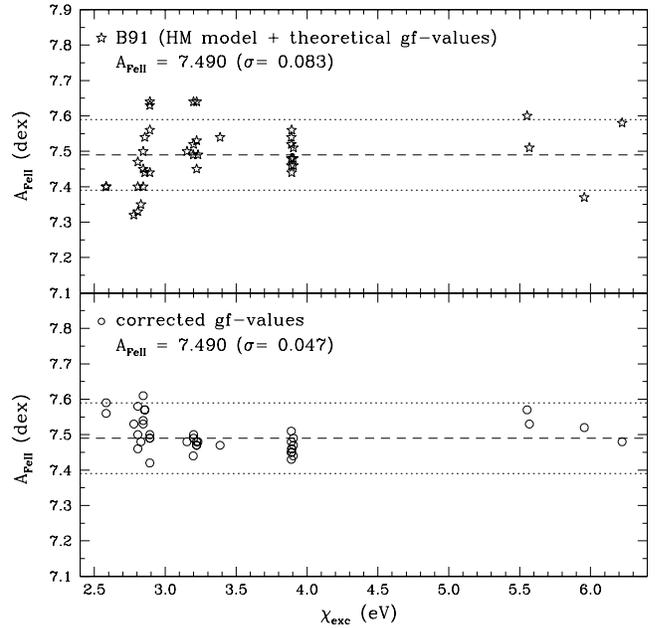}}
\caption{Iron abundances from \ion{Fe}{ii} lines obtained by Bi\'emont et al. (1991) using
theoretical $gf$-values (upper panel) and re-scaled abundances using our improved
$gf$-values (lower panel). Dashed and dotted lines are shown at the median value and 
$\pm$ 0.1 dex, respectively.}                      
\label{biemont}    
\end{figure}

\begin{table*}
\caption{Corrected literature solar iron abundances from \ion{Fe}{ii} lines using our line list}
\label{sunfe2}
\centering          
\renewcommand{\footnoterule}{}  
\begin{tabular}{cccccccccccccccccc} 
\hline\hline                
\scriptsize
{Model}  & {gf-value} & $A_{Fe}$(literature) & \# lines &  {Reference} & $A_{Fe}$(corrected) \\
\hline    
    HM & laboratory  & 7.480 ($\sigma$ = 0.084) & 15 & Hannaford et al. 1992 (H92) & 7.470 ($\sigma$ = 0.030)  \\
    HM & laboratory  & 7.420 ($\sigma$ = 0.088) & 13 & Schnabel et al. 1999 (S99)  & 7.450 ($\sigma$ = 0.054) \\
    3D & laboratory  & 7.420 ($\sigma$ = 0.099) & 15 & Asplund et al. 2000 (A00)   & 7.450 ($\sigma$ = 0.043) \\
    HM & theoretical & 7.490 ($\sigma$ = 0.083) & 39 & Biemont et al. 1991 (B91)   & 7.490 ($\sigma$ = 0.047) \\
\hline                                 
\hline                                 
\end{tabular}
\end{table*}

\begin{figure}
\resizebox{\hsize}{!}{\includegraphics{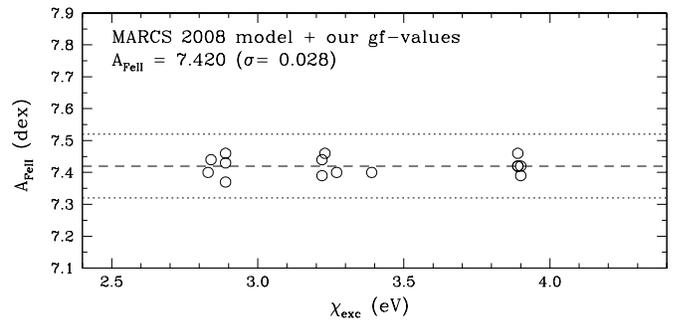}}
\caption{Iron abundances based on 15 \ion{Fe}{ii} lines with our $gf$-values and
the MARCS 2008 solar model atmosphere. Dashed and dotted lines are shown at the median value and 
$\pm$ 0.1 dex, respectively.}                      
\label{asplund}    
\end{figure}

\begin{table}
\caption{Centre-of-the-disk equivalent widths used to determine
the solar iron abundance from different model atmospheres (Table 4).
The EW are the average values presented in Holweger et al. (1990),
Bi\'emont et al. (1991), Hannaford et al. (1992), and Grevesse \& Sauval (1999).}
\label{ewfe2}
\centering 
\renewcommand{\footnoterule}{}  
\begin{tabular}{cr} 
\hline\hline                
\scriptsize
{$\lambda$ (\AA)}  & EW (m\AA)    \\
\hline    
   4576.340  &  67.0 \\
   4620.521  &  55.4 \\
   4656.981  &  35.6 \\
   5234.625  &  88.3 \\
   5264.812  &  47.5 \\
   5414.073  &  27.9 \\
   5525.125  &  12.7 \\
   5627.497  &   8.1 \\
   6432.680  &  43.4 \\
   6516.080  &  57.0 \\
   7222.394  &  20.3 \\
   7224.487  &  20.9 \\
   7449.335  &  19.4 \\
   7515.832  &  15.0 \\
   7711.724  &  50.1 \\
\hline                                 
\hline                                 
\end{tabular}
\end{table}

\begin{table}
\caption{Solar iron abundances based on 15 \ion{Fe}{ii} lines (Table 3) with our 
$gf$-values and different model atmospheres}
\label{sunfe2}
\centering 
\renewcommand{\footnoterule}{}  
\begin{tabular}{cccccccccccccccccc} 
\hline\hline                
\scriptsize
{Model}  & $v_{mic}$           & $A_{FeII}$($\sigma$) \\
{     }  & (km s$^{-1}$) & (dex)  \\
\hline    
  HM         & 1.09 & 7.46 ($\sigma$ = 0.028)   \\
Kurucz OVER  & 0.98 & 7.48 ($\sigma$ = 0.026)   \\
Kurucz NOVER & 0.91 & 7.43 ($\sigma$ = 0.027)   \\
MARCS 1997   & 0.92 & 7.44 ($\sigma$ = 0.029)   \\
MARCS 2008   & 0.91 & 7.42 ($\sigma$ = 0.028)   \\
$<$3D$>$     & 0.92 & 7.47 ($\sigma$ = 0.027)   \\
  3D         &      & 7.45 (re-scaled from A00) \\
\hline                                 
 mean        &      & 7.45 ($\sigma$ = 0.022) \\
\hline                                 
\hline                                 
\end{tabular}
\end{table}

\section{The metal-poor stars HD 140283 and HD 148816}

As shown in Sect. 3, our $gf$-values are adequate for 
solar metallicity stars. However, many \ion{Fe}{ii} lines that are
useful in metal-poor stars are blended or too strong in the Sun.
Therefore, we performed further tests of our line list using
the metal-poor stars HD 148816 and HD 140283,
which have roughly solar effective temperature, but an iron abundance 
about 5 times and 200 times lower than in the Sun, respectively.

The sample stars were observed with HIRES (Vogt et al. 1994) at the 
Keck I telescope in June 2005, i.e., after the HIRES upgrade (August 2004)
which improved the efficiency, spectral coverage and spectral resolution of HIRES. 
A resolving power of $R \approx 10^5$ was achieved using a
0.4"-wide slit. The combined spectra of HD 140283 have S/N of about 800 and 900 per pixel
at 5000 and 6500 \AA, respectively, while in the case of HD 148816 we achieved
S/N of $\approx$ 700 and 800 in the same regions. 

The superb quality of the spectra
guarantees a very stringent test of our line list, as the photon noise will not
significantly influence the line-to-line scatter in the iron abundance. A full
description of the data for these and other metal-poor stars observed for 
the determination of isotopic lithium abundances and tests of stellar parameters
and model atmospheres will be presented in 
Asplund \& Mel\'endez (2009, in preparation) and Mel\'endez et al. (2009, in preparation).

The spectra of HD 140283 and HD 148816 were scrutinized for relatively clean 
\ion{Fe}{ii} lines, resulting in a total of 27 different lines appropriate for analysis,
with 20 suitable lines available in HD 140283 and 23 in HD 148816.
In Table 5 we show the equivalent widths measured using IRAF.
The calculations were performed using the 2002 version of MOOG (Sneden 1973)
and employing Kurucz overshooting model atmospheres. 
The stellar parameters were determined following Mel\'endez et al. (2006a)
and are presented in Table 6.
Since here we are interested in the line-to-line scatter due to errors in the 
$gf$-values, the specific choice of model atmosphere and stellar parameters 
is irrelevant.
In addition to the test of our $gf$-values, we also tested the critical compilation 
of atomic transition probabilities by Fuhr \& Wiese (2006). 

The results are shown in Table 6 and Figs. 6 and 7. As can be seen, the
present line list is precise, providing a line-to-line scatter 
in the iron abundance as low as 0.04 dex in the case of HD 148816, while for
the same star the line list of Fuhr \& Wiese (2006) provides an uncomfortably
large scatter of 0.18 dex. For star HD 140283, which is much more metal-poor,
the scatter obtained with our line list is 0.01 dex higher than for
the moderately metal-poor star HD 148816, as expected due to the 
faintness of some \ion{Fe}{ii} lines in HD 140283. Our scatter of 0.05 dex
for HD 140283 is considerably lower than the scatter of 0.11 dex
obtained from the FW06 compilation. The performance of the FW06 line list
is much worse for HD 148816 ($\sigma$ = 0.18 dex) than for 
HD 140283 ($\sigma$ = 0.11 dex), while in our case the performance
was about the same ($\sigma$ = 0.04 and 0.05 dex). Thus, our line
list is almost immune to the particular choice of lines and can
be safely applied even when only a few lines are available for analysis.

\begin{table}
\caption{Equivalent widths (m\AA) for \ion{Fe}{ii} lines in HD 148816 and HD 140283.}
\label{ewhires}
\centering          
\renewcommand{\footnoterule}{}  
\begin{tabular}{crr} 
\hline\hline                
$\lambda (\AA)$  & HD 148816 & HD 140283 \\
\hline    
   4128.75 &   26.6 &  ---- \\
   4178.86 &   68.5 &  17.8 \\
   4233.17 &   ---- &  41.3 \\
   4416.83 &   61.6 &  10.3 \\
   4489.18 &   45.5 &   5.1 \\
   4491.40 &   54.8 &   7.1 \\
   4508.29 &   68.0 &  15.8 \\
   4515.34 &   58.5 &  11.6 \\
   4520.22 &   61.0 &  11.3 \\
   4522.63 &   ---- &  22.1 \\
   4541.52 &   45.3 &   5.2 \\
   4555.89 &   73.0 &  15.6 \\
   4576.34 &   45.0 &   4.5 \\
   4582.84 &   36.6 &   2.8 \\
   4583.84 &   ---- &  37.0 \\
   4731.45 &   ---- &   3.5 \\
   4923.93 &  126.5 &  57.0 \\
   5018.44 &  140.5 &  67.0 \\
   5197.58 &   62.2 &  10.7 \\
   5234.62 &   66.2 &  12.6 \\
   5264.81 &   27.3 &  ---- \\
   5284.11 &   36.4 &  ---- \\
   5414.07 &   12.0 &  ---- \\
   5425.26 &   22.1 &  ---- \\
   6369.46 &    7.4 &  ---- \\
   6432.68 &   21.8 &  ---- \\
   6516.08 &   32.0 &   2.2 \\
\hline                                 
\hline                                 
\end{tabular}
\end{table}

\begin{table}
\caption{Iron abundances from \ion{Fe}{ii} lines in the metal-poor stars HD 148816 and HD 140283.}
\label{pobres}
\centering          
\renewcommand{\footnoterule}{}  
\begin{tabular}{cccccccccccccccc} 
\hline\hline                
Star  & \tsin/log $g$/$v_{mic}$ & our line list & FW06 & n \\
      &                         & $A_{FeII}$ ($\sigma$) & $A_{FeII}$ ($\sigma$)\\
{}        &  (K/dex/km s$^{-1}$) & (dex)  & (dex) \\
\hline    
HD 148816 &  5825/4.14/1.20  & 6.72 (0.04) & 6.67 (0.18) & 23 \\
HD 140283 &  5717/3.70/1.35  & 5.10 (0.05) & 5.00 (0.11) & 20 \\
\hline                                 
\hline                                 
\end{tabular}
\end{table}

\begin{figure}
\resizebox{\hsize}{!}{\includegraphics{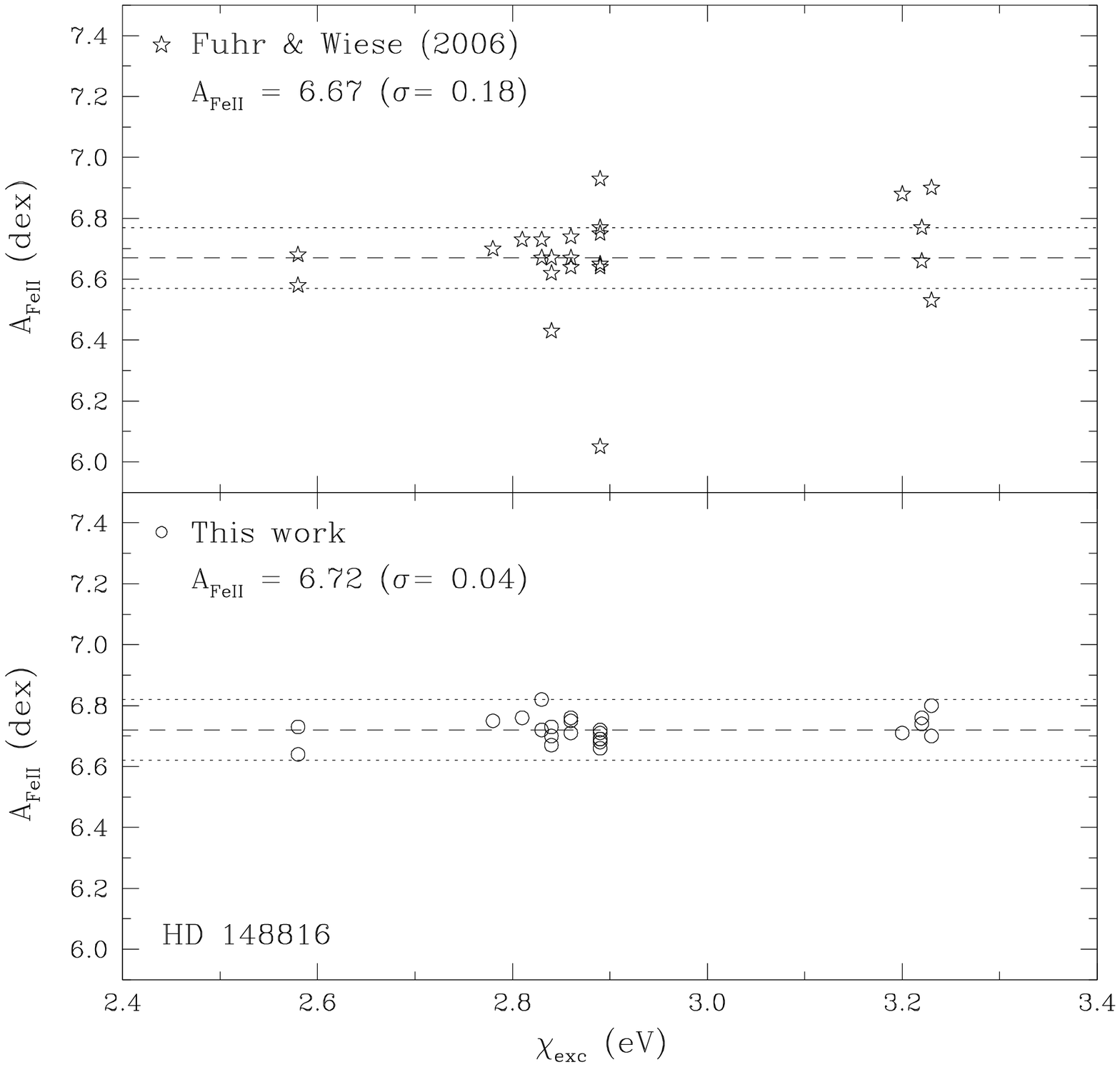}}
\caption{Iron abundances in HD 148816 from 23 \ion{Fe}{ii} lines using the 
Fuhr \& Wiese (2006) oscillator strengths (upper panel) and
our improved $gf$-values (lower panel). Dashed and dotted lines 
are shown at the median value and $\pm$ 0.1 dex, respectively.}                      
\label{hd148816}    
\end{figure}

\begin{figure}
\resizebox{\hsize}{!}{\includegraphics{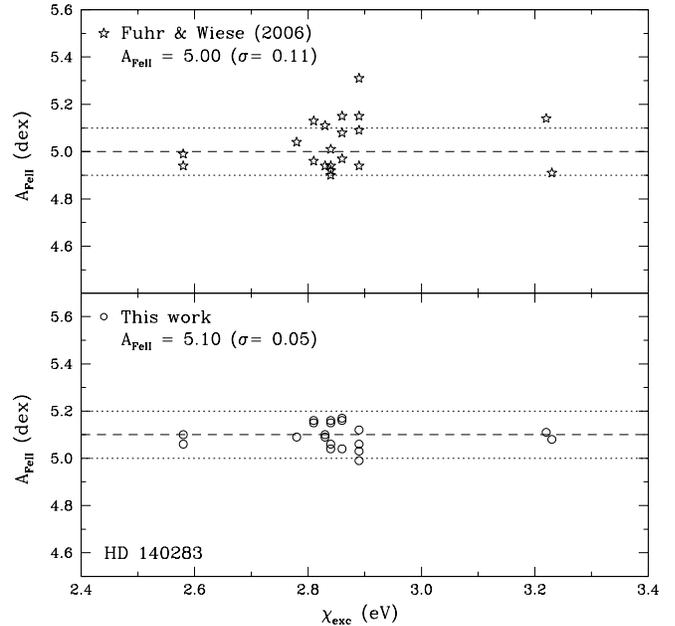}}
\caption{Iron abundances in HD 140283 from 20 \ion{Fe}{ii} lines using the 
Fuhr \& Wiese (2006) oscillator strengths (upper panel) and
our improved $gf$-values (lower panel). Dashed and dotted lines 
are shown at the median value and $\pm$ 0.1 dex, respectively.}                      
\label{hd148816}    
\end{figure}

\section{Comparison with other line lists}

In their work on RR Lyrae stars, Lambert et al. (1996)
performed a critical selection of oscillator strengths 
for \ion{Fe}{ii} lines based on a combination of
laboratory, theoretical, and solar $gf$-values.
The agreement with our work is very good, with a
mean difference of only -0.01 dex (this work - Lambert et al.)
and $\sigma$ = 0.06 dex.

We also compared our $gf$-values with other values
in the literature. This is shown in Table 7,
where the mean difference, the line-to-line scatter ($\sigma$),
and the number of lines in common are given for 18
different works in the literature. As can be seen,
the $gf$-values adopted by most works have a zero-point 
in agreement with ours within 0.02 dex, but in
some cases the standard deviation is too great
(Aoki et al. 2005; Ivans et al. 2006). In other
cases, both the mean difference and scatter are
too large (Blackwell et al. 1980; Bensby et al. 2003; 
Fran\c cois et al. 2003). 

The large difference in zero point with both Blackwell et al. (1980) 
and Gurtovenko \& Kostik (1989) stem from a much
higher solar iron abundance used by them in
their derivation of solar oscillator strengths.
The Gurtovenko \& Kostik (1989) set
of solar $gf$-values seems precise
($\sigma$ = 0.05), with even lower scatter than
the set of solar $gf$-values recently obtained by Sousa et al. (2008).

The large difference in zero point with the
$gf$-values used by Bensby et al. (2003) is
due to their use of uncorrected theoretical oscillator
strengths by Raassen \& Uylings (1998). 
The log $gf$-value of the 4993.36 \AA\ \ion{Fe}{ii} line
given in Bensby et al. is incorrectly quoted as 
-3.52. (Actually this is the value given by
Raassen \& Uylings for the 4992.47 \AA\ line.)
According to Raassen \& Uylings (1998), log $gf$ = -3.68 for the
4993.36 \AA\ line. Besides the zero-point issue, the
scatter of the $gf$-values used by
Bensby et al. is also relatively large ($\sigma$ = 0.08 dex).
Thus, the Raassen \& Uylings (1998) oscillator strengths 
are not recommended due to both their inaccuracy and imprecision.
This set of $gf$-values has been also used by
Lecureur et al. (2007) in the analysis of metal-rich
Bulge giants. Care should be taken when studying such
metal-rich cool giants, because very few unblended \ion{Fe}{ii} lines are available
for analysis, making the use of a precise line list mandatory.

The oscillator strengths by Chen et al. (2003),
Reddy et al. (2003), and Santos et al. (2004) show
the lowest scatter ($\sigma = 0.03-0.04$ dex) with
respect to our $gf$-values, but the 
Chen et al. (2003) oscillator strengths have a large zero-point
difference (0.07 dex).

\begin{table}
\caption{Mean difference between our log $gf$ values and those adopted in the literature}
\label{literature}
\centering          
\renewcommand{\footnoterule}{}  
\begin{tabular}{lrrcc} 
\hline\hline                
\scriptsize
{Reference}  & {This work - literature} & \# lines  \\
\hline    
Blackwell et al. (1980)     & 0.15 ($\sigma$ = 0.09) & 41 \\
Gurtovenko \& Kostik (1989) & 0.21 ($\sigma$ = 0.05) & 55 \\
Lambert et al. (1996) & -0.01 ($\sigma$ = 0.06) & 24 \\
Fulbright (2000)      & -0.01 ($\sigma$ = 0.07) & 26 \\
Bensby et al. (2003)  &  0.11 ($\sigma$ = 0.08) & 37 \\
Chen et al. (2003)    &  0.07 ($\sigma$ = 0.04) & 16 \\
Fran{\c c}ois et al. (2003) & -0.09 ($\sigma$ = 0.12) & 12 \\
Gratton et al. (2003) & -0.02 ($\sigma$ = 0.08) & 42 \\
Gratton et al. (2003) ($\lambda > 4600 \AA$) & 0.00 ($\sigma$ = 0.06) & 34  \\
Korn et al. (2003)    &  0.05 ($\sigma$ = 0.07) & 35 \\
Reddy et al. (2003)   & -0.01 ($\sigma$ = 0.04) &  9 \\
Sneden et al. (2003)  &  0.02 ($\sigma$ = 0.09) & 11 \\
Nissen et al. (2004)  &  0.04 ($\sigma$ = 0.06) & 19 \\
Santos et al. (2004)  &  0.01 ($\sigma$ = 0.03) & 12 \\
Aoki et al. (2005)    &  0.01 ($\sigma$ = 0.11) & 19 \\
Sadakane et al. (2005) & 0.02 ($\sigma$ = 0.06) & 11 \\
Ivans et al. (2006)   &  0.02 ($\sigma$ = 0.11) & 19 \\
Randich et al. (2006) &  0.06 ($\sigma$ = 0.06) & 10 \\ 
Sousa et al. (2008)   &  0.00 ($\sigma$ = 0.07) & 31 \\
\hline                                 
\hline                                 
\end{tabular}
\end{table}

\section{Conclusions}

We have obtained accurate and precise oscillator strengths for \ion{Fe}{ii} lines.
Our $gf$-values were tested using the Sun and the metal-poor stars HD 148816 and HD 140283, 
for which standard deviations of $\sigma$  = 0.03, 0.04, and 0.05 dex are found, respectively.
The standard error for the mean iron abundance is negligible ($\sigma_\mathrm{mean} \leq$ 0.01 dex),
and therefore the error in $gf$-values of \ion{Fe}{ii} lines is no longer a limitation
for high-precision stellar abundance work. Now the main uncertainties related to
determining iron abundances are the adopted stellar parameters, line
formation treatment, model atmospheres, blends, and errors in the measurement of
equivalent widths (or spectral synthesis fitting).

\begin{acknowledgements}
We thank the referee (Prof. D. Lambert) for his constructive comments.
This work has been partially supported by FAPESP (2005/00397-1) and CNPq (Brazil), 
and FCT (project PTDC/CTE-AST/65971/2006, and Ciencia 2007 program).
\end{acknowledgements}

\end{document}